\documentclass{article}%
\usepackage{amsfonts}
\usepackage{amsmath}
\usepackage{amssymb}
\usepackage{graphicx}%
\providecommand{\U}[1]{\protect\rule{.1in}{.1in}}

\begin{document}

\title{Geometry from Information Geometry\thanks{Presented at MaxEnt 2015, the 35th
International Workshop on Bayesian Inference and Maximum Entropy Methods in
Science and Engineering (July 19--24, 2015, Potsdam NY, USA). }}
\author{Ariel Caticha\\{\small Department of Physics, University at Albany--SUNY, Albany, NY 12222,
USA}}
\date{}
\maketitle

\begin{abstract}
We use the method of maximum entropy to model physical space as a curved
statistical manifold. It is then natural to use information geometry to
explain the geometry of space. We find that the resultant information metric
does not describe the full geometry of space but only its conformal geometry
--- the geometry up to local changes of scale. Remarkably, this is precisely
what is needed to model \textquotedblleft physical\textquotedblright\ space in
general relativity.

\end{abstract}

\section{Introduction}

The motivation behind the program of Entropic Dynamics is the realization that
the formulation of physical theories makes essential use of concepts that are
clearly designed for processing information. Prominent examples include
probability, entropy, and --- as we will argue in this work --- geometry. This
suggests that the connection between physics and nature is somewhat indirect:
the goal of physics is not to provide a direct and faithful image of nature
but to provide a framework for processing information and making inferences
\cite{Caticha 2010a}-\cite{Caticha 2015}.

This view imposes severe restrictions on physical models because the tools and
methods of physics must inevitably reflect their inferential origins.
Probabilities, for example, must necessarily be epistemic; they are, after
all, the tools that have been designed to quantify uncertainty. The entropies
must be information entropies; they are tools for updating or assigning
probabilities. What could perhaps be most surprising is that even the
geometries that pervade physics might also be of statistical origin; it might
be possible to explain them in terms of information geometry \cite{Caticha
2012}\cite{Amari 1985}.

As with any application of entropic methods, Entropic Dynamics requires that
we specify the subject matter (the microstates) and the relevant information
(the constraints) on the basis of which we will carry out our inferences. In
this paper we take the first step towards formulating an entropic dynamics of
gravity --- we identify the subject matter. (Two other relevant contributions
in this direction are \cite{Ipek Caticha 2015}\cite{Nawaz et al 2015}.)

We use the method of maximum entropy to model \textquotedblleft
physical\textquotedblright\ three dimensional space as a curved statistical
manifold. The basic idea is that the points of space are not defined with
perfect resolution; they are not structureless dots. When we say that a
particle is located at a point $x$, its actual true location $x^{\prime}$ is
uncertain and lies somewhere in the vicinity of $x$. Thus, to each point $x$
in space one associates a probability distribution, $p(x^{\prime}|x)$.\ In
this model space is a statistical manifold and is automatically endowed with
an information metric. It is important to emphasize that information geometry
yields positive definite metrics which apply to Riemannian manifolds. The
problem of modelling the pseudo-Riemannian geometry of spacetime remains open.

We find that the resultant information geometry does not specify the full
geometry of the statistical manifold. It allows an arbitrary choice of the
local scale which means that what is described is the conformal geometry of
space. Remarkably, this is precisely what is needed to model \textquotedblleft
physical\textquotedblright\ space in general relativity. Indeed, there is
convincing evidence that the dynamical degrees of freedom of the gravitational
field represent the conformal geometry of space-like hypersurfaces embedded in
space-time \cite{York 1972}-\cite{Gomes et al 2011}.

The construction is straightforward except for one technical difficulty. Since
coordinates do not themselves carry any information --- we can always change
coordinates --- it is essential to maintain covariance under coordinate
transformations. The difficulty is that the expected value constraints
required by the method of maximum entropy do not transform covariantly. The
problem is overcome by applying the method of maximum entropy in the flat
tangent spaces and using the exponential map to induce probabilities on the
manifold itself.

In our brief closing remarks we observe that when space is modeled as a
statistical manifold it acquires a curious hybrid character and exhibits
features that are typical of discrete spaces while maintaining the
calculational advantages of continuous manifolds. Finally, we show that the
information volume of a region of space is a measure of the effective number
of distinguishable points and is also a measure of its entropy.

\section{Distinguishability and information distance}

Our subject is space which we model as a smooth three-dimensional manifold
$\mathbf{X}$. There are no external rulers and therefore there is no
externally imposed notion of distance. The main assumption is that
$\mathbf{X}$ is blurred; its points are defined with a finite resolution.
Consider a test particle (or a field variable) located at $x\in\mathbf{X}$
with coordinates $x^{a}$, $a=1,2,3$. When we say that the test particle is at
$x$ it turns out that it is actually located at some unknown $x^{\prime}$
somewhere in its vicinity. The uncertainty in $x^{\prime}$ is represented by a
probability distribution, $p(x^{\prime}|x)$. (The probability that $x^{\prime
}$ lies within $d^{3}x^{\prime}$ is $p(x^{\prime}|x)d^{3}x^{\prime}$.) We need
not, at this point, specify the physical origin of the uncertainty. Since to
each point $x\in\mathbf{X}$ one associates a probability distribution
$p(x^{\prime}|x)$ the space $\mathbf{X}$ is a special type of statistical manifold.

In a generic statistical manifold $\mathbf{M}$ one associates a probability
distribution $p(\xi|x)$ to each point $x\in\mathbf{M}$. The variables $\xi$
and $x$ need not represent physical quantities of the same kind. For example,
in the case of a gas $\xi$ can represent the positions and momenta of the
molecules, while $x$ could stand for the temperature and volume of the gas.
Here we deal with a special type of statistical manifold in which both
$x^{\prime}$ and $x$ are positions in the same space.

Coordinates $x^{a}$ are introduced to distinguish one point from another, but
if points are blurred we cannot fully distinguish the point at $x^{a}$ from
another point at $x^{a}+dx^{a}$. We seek a quantitative measure of the extent
to which these two distributions can be distinguished. It is remarkable that
this measure is determined uniquely by imposing certain symmetries that are
natural for statistical manifolds --- the invariance under Markovian
embeddings. It is even more remarkable that this unique measure has all the
properties of a distance \cite{Caticha 2012}\cite{Amari 1985}. Such
\emph{information distance} is given by
\begin{equation}
d\ell^{2}=g_{ab}\,(x)dx^{{}a}dx^{{}b}\,\,, \label{info metric a}%
\end{equation}
where the metric tensor $g_{ab}$ --- \emph{the information metric} --- is
given by,
\begin{equation}
g_{ab}\,(x)=\int dx^{\prime}\,p(x^{\prime}|x)\,\partial_{a}\log p(x^{\prime
}|x)\,\partial_{b}\log p(x^{\prime}|x)~.\,\, \label{info metric b}%
\end{equation}
(We adopt the standard notation $\partial_{a}=\partial/\partial x^{a}$ and
$dx^{\prime}=d^{3}x^{\prime}$.)

To complete the definition of the information geometry of $\mathbf{X}$ we must
specify a connection or covariant derivative $\nabla$. This allows us to
introduce notions of parallel transport, curvature, and so on. Although we
will not use covariant derivatives in this work, it appears that the
Levi-Civita connection, defined so that $\nabla_{a}g_{bc}=0$, is the candidate
of choice. It is the simplest among all the $\alpha$-connections \cite{Amari
1985}, and it is also the most natural because it does not require that any
additional structure be imposed on the Hilbert space of functions $(p)^{1/2}$
\cite{Brody Hughston 1997}.

Thus the space $\mathbf{X}$ inherits its geometry from the family of
distributions $p(x^{\prime}|x)$. Even at this early stage we can envision
potentially important consequences for later applications to quantum gravity.
Contrary to naive expectation the statistical manifolds proposed here are not
rougher, more irregular, than those needed to describe classical gravity. In
fact they may be considerably smoother because irregularities at scales
smaller than the local uncertainty become meaningless.

\section{Using maximum entropy to assign $p(x^{\prime}|x)$}

Next we use the method of maximum entropy to assign the distribution
$p(x^{\prime}|x)$. The central physical assumption is that the physically
relevant information that is necessary to model space is captured by
constraints on the expectation of $x^{\prime}$ and of its uncertainty.
Therefore one is led to consider the expected value
\begin{equation}
\langle x^{\prime}{}^{a}\rangle=%
{\textstyle\int}
dx^{\prime}\,p(x^{\prime}|x)\,x^{\prime}{}^{a}~,
\end{equation}
and the variance-covariance matrix
\begin{equation}
\left\langle (x^{\prime a}-x^{a})(x^{\prime}{}^{b}-x^{b})\right\rangle =%
{\textstyle\int}
dx^{\prime}\,p(x^{\prime}|x)(x^{\prime}{}^{a}-x^{a})(x^{\prime}{}^{b}-x^{b})~.
\end{equation}

\paragraph*{The problem with covariance}

A problem arises immediately because in a curved space neither of these
constraints is covariant. To see the difficulty consider a change of
coordinates. Let $x^{i}=X^{i}(x^{a})$ and $x^{\prime i}=X^{i}(x^{\prime a})$
where the indices $abc$ and $ijk$ denote old and new coordinates respectively.
Taylor expand $x^{\prime i}$ in $(x^{\prime a}-x^{a})$,
\begin{equation}
x^{\prime i}-x^{i}=X_{a}^{i}\left(  x^{\prime a}-x^{a}\right)  +\frac{1}%
{2}X_{ab}^{i}\left(  x^{\prime a}-x^{a}\right)  \left(  x^{\prime b}%
-x^{b}\right)  +\ldots\label{non cov a}%
\end{equation}
where
\begin{equation}
X_{a}^{i}=\partial_{a}X^{i}(x)=\frac{\partial x^{i}}{\partial x^{a}}%
\quad\text{and}\quad X_{ab}^{i}=\partial_{a}\partial_{b}X^{i}(x)=\frac
{\partial^{2}x^{i}}{\partial x^{a}\partial x^{b}}~.
\end{equation}
Taking the expected value with the scalar $dx^{\prime}p(x^{\prime}|x)$ gives
\begin{equation}
\left\langle x^{\prime i}-x^{i}\right\rangle =X_{a}^{i}\left\langle x^{\prime
a}-x^{a}\right\rangle +\frac{1}{2}X_{ab}^{i}\left\langle \left(  x^{\prime
a}-x^{a}\right)  \left(  x^{\prime b}-x^{b}\right)  \right\rangle
+\ldots\label{non cov b}%
\end{equation}
which shows that we can impose $\left\langle x^{\prime a}\right\rangle =x^{a}%
$, but on changing coordinates we will have
\begin{equation}
\left\langle x^{\prime i}\right\rangle =x^{i}+O(\sigma^{2})
\end{equation}
where $O(\sigma^{2})$ represents a non-negligible correction of the order of
the width of the distribution. Therefore neither coordinate differences
$x^{\prime a}-x^{a}$, eq.(\ref{non cov a}), nor their expected values,
eq.(\ref{non cov b}), transform as components of a vector.

\paragraph*{The Exponential map}

The problem with the noncovariance of expected values can be traced to the
fact that the statistical manifold $\mathbf{X}$ is curved. To evade this
problem the entropy maximization will be carried out in the flat spaces
$\mathbf{T}_{P}$ that are tangent to $\mathbf{X}$ at each point $P$ and then a
special map --- the \emph{exponential map} --- will be used to obtain the
corresponding induced distributions on $\mathbf{X}$.

The exponential map is defined as follows. Assume that the manifold
$\mathbf{X}$ has a metric $g_{ab}$ --- we shall later see that it does. Then
we can construct geodesics. Consider the space $\mathbf{T}_{P}$ that is
tangent to $\mathbf{X}$ at $P$. Each vector $\vec{y}\in\mathbf{T}_{P}$ defines
a unique geodesic through $P$. Let the points $Q\in\mathbf{X}$ on the geodesic
through $P$ with tangent vector $\vec{y}$ be denoted by $Q(\vec{y};\lambda)$
where $\lambda$ is an affine parameter such that $Q(\vec{y};\lambda=0)=P$.
Then the point with affine parameter $\lambda=1$ is assigned coordinates
$y^{i}$,
\begin{equation}
Q^{i}(\vec{y};\lambda=1)=y^{i}~. \label{RNC}%
\end{equation}
This construction maps a straight line in the flat tangent space
$\mathbf{T}_{P}$ to a straight line in the curved manifold $\mathbf{X}$. The
set of vectors $\lambda\vec{y}\in\mathbf{T}_{P}$ is mapped to the set of
points $x^{i}=\lambda y^{i}\in\mathbf{X}$. The map $\mathbf{T}_{P}%
\rightarrow\mathbf{X}$ is called \emph{the exponential map} and the
corresponding coordinates have several useful properties \cite{Sternberg
2012}. For our purposes we will only need
\begin{equation}
g_{ij}(P)=\delta_{ij}\quad\text{and}\quad\partial_{k}g_{ij}(P)=0~.
\end{equation}
These coordinates are called Riemann Normal Coordinates at $P$ (denoted
NC$_{P}$).\footnote{The exponential map from $\mathbf{T}_{P}$ to $\mathbf{X}$
is 1-1 only within some neghborhood of $P$ --- the so-called \emph{normal}
neighborhood. Beyond this neighborhood the geodesics in a curved manifold may
cross. In such cases the mapping remains well defined but it no longer serves
the useful purpose of defining a coordinate system. For the smooth statistical
manifolds that interest us we can expect that the normal neighborhoods will
extend over rather large regions which renders the exponential map
particularly useful.}

\paragraph*{Using MaxEnt on the tangent spaces}

Our goal is to assign the distribution $p(Q|P)=p(x^{\prime}|x)$ on the basis
of information about the expected position and its uncertainty. This
information is provided through constraints defined on the tangent space
$\mathbf{T}_{P}$. We use the method of maximum entropy to assign a
distribution $\hat{p}(\vec{y}|P)$ on $\mathbf{T}_{P}$ and the exponential map
is then used to induce the corresponding distribution $p(x^{\prime}|x)$ on
$\mathbf{X}$.

Consider a point $P\in\mathbf{X}$ with generic coordinates $x^{a}$ and a
positive definite tensor field $\gamma^{ab}(x)$. The components of a vector
$\vec{y}\in\mathbf{T}_{P}$ are $y^{a}$. The distribution $\hat{p}(\vec{y}|P)$
is assigned on the basis of information about the expected location on
$\mathbf{T}_{P}$,
\begin{equation}
\langle y^{a}\rangle_{P}=%
{\textstyle\int}
d^{3}y\,\hat{p}(y|P)\,y^{a}=0~, \label{exp y}%
\end{equation}
and the variance-covariance matrix
\begin{equation}
\left\langle y^{a}y^{b}\right\rangle _{P}=%
{\textstyle\int}
d^{3}y\,\hat{p}(y|P)y^{a}y^{b}=\gamma^{ab}(P)~. \label{exp yy}%
\end{equation}
It is always possible to transform to new coordinates
\begin{equation}
x^{i}=X^{i}(x^{a})\quad\text{such that\quad}\gamma^{ij}(P)=\delta^{ij}%
\quad\text{and}\quad\partial_{k}\gamma^{ij}(P)=0~. \label{NC a}%
\end{equation}
(If $\gamma^{ab}$ where a metric tensor this would be a transformation to
NC$_{P}$.) The new components of $\vec{y}$ are
\begin{equation}
y^{i}=X_{a}^{i}y^{a}\quad\text{where}\quad X_{a}^{i}=\partial_{a}%
X^{i}(x)=\frac{\partial x^{i}}{\partial x^{a}}~, \label{NC b}%
\end{equation}
and the constraints (\ref{exp y}) and \ref{exp yy} take a simpler form,
\begin{equation}
\langle y^{i}\rangle_{P}=0\quad\text{and}\quad\left\langle y^{i}%
y^{j}\right\rangle _{P}=\delta^{ij}~. \label{constr}%
\end{equation}

The distribution we seek is that which maximizes the entropy
\begin{equation}
S[\hat{p},q]=-\int dy\,\hat{p}(\vec{y}|P)\log\frac{\hat{p}(\vec{y}|P)}%
{q(\vec{y})}\,
\end{equation}
relative to a measure $q(\vec{y})$. Since $\mathbf{T}_{P}$ is flat we take
$q(\vec{y})$ to be constant and we may ignore it. Maximizing $S[\hat{p},q]$
subject to the constraints (\ref{constr}) and normalization yields
\begin{equation}
\hat{p}(\vec{y}|P)=\exp\left[  -\alpha-\beta_{i}y^{i}-\frac{1}{2}\gamma
_{ij}y^{i}y^{j}\right]  ~,
\end{equation}
where $\alpha$, $\beta_{i}$, and $\gamma_{ij}$ are Lagrange multipliers.
Requiring that $\hat{p}(\vec{y}|P)$ satisfy the constraints implies that
$e^{-\alpha}$ is just a normalization constant, that the three multipliers
$\beta_{i}$ vanish, and that the matrix $\gamma_{ij}$ is the inverse of the
covariance matrix $\gamma^{ij}=\delta^{ij}$, that is, $\gamma_{ij}\gamma
^{jk}=\delta_{i}^{k}$. Therefore $\gamma_{ij}=\delta_{ij}$ and
\begin{equation}
\hat{p}(\vec{y}|P)=\frac{(\det\gamma_{ij})^{1/2}}{(2\pi)^{3/2}}\,\exp\left[
-\frac{1}{2}\gamma_{ij}y^{i}y^{j}\right]  =\frac{1}{(2\pi)^{3/2}}\,\exp\left[
-\frac{1}{2}\delta_{ij}y^{i}y^{j}\right]  ~. \label{NC Gaussian a}%
\end{equation}
We can now transform back to the original coordinates $y^{a}$ using the
inverse of eq.(\ref{NC b}),
\begin{equation}
y^{a}=X_{i}^{a}y^{i}\quad\text{and}\quad\gamma_{ab}=X_{a}^{i}X_{b}^{j}%
\delta_{ij}~. \label{gamma ab}%
\end{equation}
The resulting distribution is also Gaussian,
\begin{equation}
\hat{p}(\vec{y}|P)=\frac{(\det\gamma_{ab})^{1/2}}{(2\pi)^{3/2}}\,\exp\left[
-\frac{1}{2}\gamma_{ab}y^{a}y^{b}\right]  ~. \label{GC Gaussian a}%
\end{equation}

Next we use an exponential map to induce the corresponding distribution
$p(x^{\prime}|x)$ on the manifold $\mathbf{X}$. We use the tensor $\gamma
^{ab}$ as if it were the inverse of a metric tensor. This allows us to define
the corresponding \textquotedblleft geodesics\textquotedblright\ and
exponential map. Let the coordinates of the point $P\in\mathbf{X}$ be denoted
$x^{i}(P)$, then the normal coordinates of neighboring points will be denoted
\begin{equation}
x^{\prime i}=x^{i}(P)+y^{i}~, \label{NC x'}%
\end{equation}
and the distribution $p(x^{\prime}|P)=p(x^{\prime i}|x^{i})$ induced by
$\hat{p}(\vec{y}|P)$, eq.(\ref{NC Gaussian a}), is
\begin{align}
p(x^{\prime i}|x^{i})  &  =\frac{(\det\gamma_{ij})^{1/2}}{(2\pi)^{3/2}}%
\,\exp\left[  -\frac{1}{2}\gamma_{ij}(x^{\prime i}-x^{i})(x^{\prime j}%
-x^{j})\right] \nonumber\\
&  =\frac{1}{(2\pi)^{3/2}}\,\exp\left[  -\frac{1}{2}\delta_{ij}(x^{\prime
i}-x^{i})(x^{\prime j}-x^{j})\right]  ~. \label{NC Gaussian b}%
\end{align}
In NC$_{P}$ the distribution (\ref{NC Gaussian b}) retains the Gaussian form,
just like (\ref{NC Gaussian a}). We can now transform back to the generic
frame $x^{a}$ of coordinates and define $p(x^{\prime a}|x^{a})$ by%
\begin{equation}
p(x^{\prime a}|x^{a})d^{3}x^{\prime a}=p(x^{\prime i}|x^{i})d^{3}x^{\prime
i}~, \label{px'x}%
\end{equation}
which is a covariant identity between scalars and holds in all coordinate
systems. In the $x^{a}$ coordinates the distribution $p(x^{\prime a}|x^{a})$
will not, in general, be Gaussian,
\begin{equation}
p(x^{\prime a}|x^{a})=\frac{(\det\gamma_{ab})^{1/2}}{(2\pi)^{3/2}}%
\,\exp\left[  -\frac{1}{2}\delta_{ij}\left(  X^{i}(x^{\prime a})-X^{i}%
(x^{a})\right)  \left(  X^{j}(x^{\prime a})-X^{j}(x^{a})\right)  \right]  ~.
\label{GC Gaussian b}%
\end{equation}

\section{The information geometry of space}

Next we calculate the information metric $g_{ab}$ associated with the
distributions (\ref{GC Gaussian b}). The direct substitution of
eq.(\ref{GC Gaussian b}) into eq.(\ref{info metric b}) yields an integral that
can be handled by transforming to NC$_{P}$. Using eq.(\ref{px'x}) and
$\partial_{a}=X_{a}^{i}\partial_{i}$, we get
\begin{equation}
g_{ab}\,(x)=X_{a}^{i}X_{b}^{j}\int d^{3}x^{\prime i}\,p(x^{\prime i}%
|x^{i})\,\partial_{i}\log p(x^{\prime i}|x^{i})\,\partial_{j}\log p(x^{\prime
i}|x^{i})~.\,\,
\end{equation}
Since $p(x^{\prime i}|x^{i})$ is Gaussian, eq.(\ref{NC Gaussian b}), this
integral is straightforward. First substitute eq.(\ref{NC x'}), then integrate
using eq.(\ref{constr}), and finally transform back to the original generic
coordinates $x^{a}$ using eq.(\ref{gamma ab}), to get
\begin{equation}
g_{ab}=X_{a}^{i}X_{b}^{j}\delta_{ij}=\gamma_{ab}~. \label{GC metric}%
\end{equation}

This result is deceptively simple:\ \emph{the information metric }$g_{ab}%
$\emph{ is the inverse of the covariance tensor }$\gamma^{ab}$\emph{ that
describes the blurriness of space}. But one should not let formal simplicity
stand in the way of appreciating its significance. The metric (\ref{GC metric}%
) represents a potentially fruitful conceptual development. The idea might
best be conveyed through a historical analogy. The concept of temperature was
first introduced as an unexplained \textquotedblleft degree of
hotness\textquotedblright. Temperature was operationally defined as whatever
was measured by peculiar devices called thermometers and eventually it came to
be interpreted as an average kinetic energy per degree of freedom. It took a
long time before arriving at the modern entropic interpretation of temperature
$T$ as a Lagrange multiplier ($\beta=1/kT$) in a maximum entropy distribution.
It is conceivable that the notion of distance might undergo a similar
development. Distance has long been taken for granted --- an unexplained
quantity measured by peculiar devices called rulers. The main result of this
paper is to suggest that the metric of space $g_{ab}$ is a statistical concept
that measures a \textquotedblleft degree of
distinguishability\textquotedblright\ and that it can be traced to Lagrange
multipliers $\gamma_{ab}$ that describe the blurriness $\gamma^{ab}$ of space .

\section{Discussion and conclusions}

\paragraph*{Canonical quantization of gravity?}

From the perspective of information geometry any attempt to quantize gravity
by imposing commutation relations on the metric tensor $g_{ab}$, that is, on
the Lagrange multipliers $\gamma_{ab}$, is misguided --- it leads to a dead
end. It would appear to be just as misguided as attempting to formulate a
quantum theory of fluids by imposing commutation relations on those Lagrange
multipliers (like temperature, pressure, or chemical potential) that define
the statistical macrostate.

\paragraph*{Dimensionless distance?}

There is one very peculiar feature of the information distance $d\ell$ in
eq.(\ref{info metric a}) that turns out to be very significant: $d\ell$ is
dimensionless. Indeed, we can easily verify in eq.(\ref{info metric b}) that
if $dx^{i}$ has units of length, then $p(x^{\prime}|x)$ has units of inverse
volume, and $g_{ij}$ has units of inverse length squared. Distances are
supposed to be measured in units of length; what sort of distance is this
$d\ell$?

A simple example will help clarify this issue. Consider two neighboring
Gaussian distributions,
\begin{equation}
p(x^{\prime}|x)=\frac{(\det\gamma_{ij})^{1/2}}{(2\pi)^{3/2}}\,\exp\left[
-\frac{1}{2}\gamma_{ij}(x^{\prime i}-x^{i})(x^{\prime j}-x^{j})\right]
\end{equation}
and $p(x^{\prime}|x+dx)$, with means $x$ and $x+dx$ and the same covariance
matrix, $\gamma_{ij}=\delta_{ij}/\sigma^{2}$. The distance between them,
eq.(\ref{info metric a}), is%
\begin{equation}
d\ell^{2}=\frac{1}{\sigma^{2}}\delta_{ij}dx^{i}dx^{j}~.
\end{equation}
This is the Euclidean metric $\delta_{ij}$ rescaled by $\sigma^{2}$. Therefore
the dimensionless $d\ell$ represents a distance measured in units of the
uncertainty $\sigma$. More generally,\emph{ the information metric }%
$g_{ij}(x)$\emph{ measures distinguishability in units of the local
uncertainty implied by the distribution }$p(x^{\prime}|x)$.

As long as we are concerned with quantifying \emph{distinguishability} no
global unit of length is needed, but if what we want is a measure of
absolute\ \emph{distance} several related questions arise: How do we compare
the uncertainties at two different locations? Or, alternatively, does the
local uncertainty provide us with a universal, global standard of length? If
it does not, and the uncertainty varies from point to point, how do we compare
lengths at different places?

The answer is that an absolute comparison of how the uncertainty varies from
point to point is objectively meaningless because there are no external
rulers. Information geometry does not provide comparisons of lengths at
different locations but it does allow us to compare short lengths at the same
location. This means that \emph{what the information geometry describes is the
conformal geometry of space}. It describes the local \textquotedblleft
shape\textquotedblright\ of space but not its absolute local \textquotedblleft
size\textquotedblright.

Nevertheless, we humans can still adopt some criterion that determines a local
scale and allows us to define length as a tool for reasoning, as an aid for
constructing pictures and models of the world. In such models geometry is
described by a metric tensor $\bar{g}_{ab}$ that is conformally related to the
information metric in (\ref{GC metric}),
\begin{equation}
\bar{g}_{ab}(x)=\sigma^{2}(x)g_{ab}(x)~.
\end{equation}
The choice of the scale factor $\sigma^{2}(x)$, which amounts to a choice of
\textquotedblleft gauge\textquotedblright, is a matter of convenience. It is
dictated by purely pragmatic considerations: \emph{length is defined so that
physics looks simple}. The scale of distance --- just like the duration of
time --- turns out to be a property not of the world but of the models we
employ to describe it. One possible choice of gauge would be, for example, to
legislate that $\sigma^{2}(x)=\sigma_{0}^{2}$ is a constant. Another
possibility is to choose $\sigma^{2}(x)$ so that the evolving
three-dimensional manifold $\mathbf{X}$ generates a four-dimensional
space-time.\footnote{The conditions for such a \emph{space-time gauge} involve
scale factors that satisfy the Lichnerowicz-York equation \cite{York
1972}\cite{OMurchadha York 1974}. Similar notions have also been proposed in
the context of Machian relational dynamics by Barbour and his collaborators.
(See \emph{e.g. }\cite{Barbour et al 2010}\cite{Gomes et al 2011}\emph{.})}

\paragraph*{The statistical state of space}

The state of space is the joint distribution of all the $y_{x}$ variables
associated to every point $x$. We assume that the $y_{x}$ variables at $x$ are
independent of the $y_{x^{\prime}}$ variables at $x^{\prime}$, and therefore
their joint distribution is a product,
\begin{equation}
\hat{P}[y|g]=%
{\textstyle\prod\limits_{x}}
\hat{p}\left(  y_{x}|x,g_{x}\right)  ~, \label{state space a}%
\end{equation}
where $g_{x}$ is short for $g_{ab}(x)$. Given $g_{ab}(x)$ at any point $x$ the
distribution $\hat{p}\left(  y_{x}|x,g_{x}\right)  $ in the tangent space
$\mathbf{T}_{x}$ is Gaussian,
\begin{equation}
\hat{p}(y_{x}|x,g_{x})=\frac{(\det g_{x})^{1/2}}{(2\pi)^{3/2}}\,\exp\left[
-\frac{1}{2}g_{ab}(x)y_{x}^{a}y_{x}^{b}\right]  ~. \label{state space b}%
\end{equation}
We conclude that the information metric $g_{ab}$ determines the statistical
state of space.

\paragraph*{Continuous and/or discrete?}

A perhaps unexpected consequence of the notion of an information distance is
the following. Suppose we want to measure the size of a finite region of space
by counting the number of points within it. Counting points depends on a
decision what we mean by a point and, in particular, on what we mean by two
different points. If we agree that two points ought to be counted separately
only when we can distinguish them then one can assert that the number of
distinguishable points in any finite region is finite. Therefore, the answer
to the old question of whether space is continuous or discrete is that, in a
certain sense, it is both. If the local uncertainty at $x$ is described as
$\sigma(x)$ then, roughly, a volume contains one distinguishable point per
volume $\frac{4}{3}\pi\sigma^{3}(x)$ and a surface contains one
distinguishable point per area $\pi\sigma^{2}(x)$. Remarkably this allows us
to compare the sizes of regions of different dimensionality: it is meaningful
to assert that a surface and a volume are of the same (information)
size\ whenever they contain the same number of distinguishable points.
Furthermore, since the number of distinguishable points within $d^{3}x$ is
$g^{1/2}d^{3}x$, this suggests that sums over distinguishable points can be
given a continuum representation by replacing sums by integrals,
\begin{equation}%
{\displaystyle\sum\nolimits_{x}}
\left(  \cdots\right)  \;\rightarrow\;\int d^{3}x\,g^{1/2}\,\left(
\cdots\right)  ~\,.
\end{equation}
It is also to be expected that modelling space as a statistical manifold will
provide a natural regulator that will eliminate the divergences that normally
afflict relativistic quantum field theories.

\paragraph*{The entropy of space}

As an example we calculate the total entropy of space,
\begin{equation}
S[\hat{P},\hat{Q}]=-\int Dy\,\hat{P}[y|g]\log\frac{\hat{P}[y|g]}{\hat{Q}%
[y|g]}\overset{\text{def}}{=}S[g] \label{Sg a}%
\end{equation}
relative to the uniform distribution%
\begin{equation}
\hat{Q}[y|g]=%
{\textstyle\prod\nolimits_{x}}
g^{1/2}(x)~, \label{prior y}%
\end{equation}
which is independent of $y$ --- a constant. Since the $y$'s in
eq.(\ref{state space a}) are independent variables the entropy is additive,
$S[g]=%
{\textstyle\sum\nolimits_{x}}
S(x)$, and we need to calculate the entropy $S(x)$ associated to a point at a
generic location $x$,
\begin{equation}
S(x)=-\int d^{3}y\,\hat{p}(y|x,g_{x})\log\frac{\hat{p}(y|x,g_{x})}{g^{1/2}%
(x)}=\frac{3}{2}\log2\pi e=s_{0}~.
\end{equation}
Thus, the entropy per point is a numerical constant ($s_{0}\approx4.2568$) and
the entropy of any region $R$ of space, $S_{R}[g]$, is just its information volume,%

\begin{equation}
S_{R}[g]=%
{\displaystyle\sum\nolimits_{x\in R}}
S(x)=s_{0}\int_{R}d^{3}x\,g^{1/2}(x)~. \label{Sg b}%
\end{equation}
Thus, the information volume $%
{\textstyle\int_{R}}
d^{3}x\,g^{1/2}$ of a region $R$ is a measure of the effective number of
distinguishable points within it and also a measure of its entropy.

\paragraph*{Summary}

Physical\ space is modelled as a statistical manifold $\mathbf{X}$ --- to each
point $x\in\mathbf{X}$ one associates a probability distribution $p(x^{\prime
}|x)$. This automatically endows the space $\mathbf{X}$ with a geometry --- an
information geometry that determines its conformal geometry.

The problem of assigning $p(x^{\prime}|x)$ in a way that guarantees covariance
is addressed by focusing attention on vector variables $y_{x}$ that live in
each tangent space $\mathbf{T}_{x}$. The method of maximum entropy is used to
assign the distributions $\hat{p}\left(  y_{x}|x\right)  $ at each $x$ and
then the exponential map $\mathbf{T}_{P}\rightarrow\mathbf{X}$ is used to
induce the corresponding distributions $p(x^{\prime}|x)$ on $\mathbf{X}$. The
validity of the construction rests on the assumption that the normal
neighborhood of every point $x$ --- the region about $x$ where the exponential
map is 1-1 --- is sufficiently large. The assumption is motivated by the
intuition that the statistical manifolds $\mathbf{X}$ are very smooth. Indeed,
when points are separated by distances less than the local uncertainty
$\sigma$ they cannot be effectively distinguished and it is not possible to
have curvatures larger than $1/\sigma$.

\paragraph*{Acknowledgments}

I would like to thank D. Bartolomeo, C. Cafaro, N. Caticha, S. DiFranzo, A.
Giffin, P. Goyal, S. Ipek, D.T. Johnson, K. Knuth, S. Nawaz, M. Reginatto, C.
Rodr\'{\i}guez, J. Skilling, and C.-Y. Tseng for many discussions on entropy,
inference, and information geometry.

\end{document}